\documentclass[twocolumn,prb,aps,draft,a4paper,nobibnotes] {revtex4}
\usepackage[final]{graphics}

\def\tessera{{\it tessera}}
\def\tesserae{{\it tesserae}}
\def\mosaico{Mosaico}
\def\order{${\cal O}$}
\def\peo{H(CH$_2$OCH$_2$)$_m$H}
\def\etal{{\it et al.}}
\def\abinitio{{\it ab initio}}
\def\cmm1{cm$^{-1}$}

\def\CeClvi-iii{(CeCl$_6$)$^{3-}$}
\def\PaClvi-ii{(PaCl$_6$)$^{2-}$}
\def\UClvi-iii{(UCl$_6$)$^{3-}$}
\def\VClvi-iii{(VCl$_6$)$^{3-}$}

\begin{document}

   \title{ 
           Parallel, linear-scaling building-block and embedding method
           based on localized orbitals and orbital-specific basis sets.
         }
   \author{Luis Seijo}
   \email[e-mail ]{luis.seijo@uam.es}
   \author{Zoila Barandiar\'an}
   \email[e-mail ]{zoila.barandiaran@uam.es}
   \affiliation{
           Departamento de Qu\'{\i}mica, C-XIV,
           Universidad Aut\'onoma de Madrid, 28049 Madrid, Spain
           \\ and \\
           Instituto Universitario de Ciencia de Materiales Nicol\'as Cabrera,
           Universidad Aut\'onoma de Madrid, 28049 Madrid, Spain
           }
   \begin{abstract}
We present a new linear scaling method for the energy minimization step
of semiempirical and first-principles Hartree-Fock and Kohn-Sham calculations. 
It is based on the self-consistent calculation of the optimum
localized orbitals of any localization method of choice
and on the use of orbital-specific basis sets.
The full set of localized orbitals of a large molecule is seen as an
orbital mosaic
where each \tessera\ is made of only a few of them.
The orbital \tesserae\ are computed out of
a set of embedded cluster pseudoeigenvalue coupled equations which are solved
in a building-block self-consistent fashion.
In each iteration, the embedded cluster equations are solved
independently of each other and, as a result,
the method is parallel at a high level of the calculation.
In addition to full system calculations, the method enables to perform
simpler, much less demanding embedded cluster calculations,
where only a fraction of the localized molecular orbitals are variational
while the rest are frozen,
taking advantage of the transferability of the localized orbitals
of a given localization method between similar molecules.
Monitoring single point energy calculations 
of large poly(ethylene oxide) molecules and 
three dimensional carbon monoxide clusters
using an extended H\"uckel Hamiltonian are presented.
   \end{abstract}
   \maketitle
\section{INTRODUCTION}
\label{SEC:Introduction}

The development of linear scaling computational methods for
electronic structure calculations in molecules and solids with a
very large number of atoms, (i.e. methods whose computational
demands grow as the first power of the size of the system,) has been
a very active and successful field in the last
decade.\cite{GOEDECKER:99,ORDEJON:00}
Linear scaling and low order scaling techniques exist for the computation of
the one-electron effective Hamiltonian matrix in first-principles calculations
(with density functional theory and wave function based 
methods),~\cite{WHITE:94,WHITE:96,ORDEJON:96,STRAIN:96,SCUSERIA:99,OCHSENFELD:98}
as well as for the energy minimization
step,~\cite{GOEDECKER:99,ORDEJON:00,YANG:91,YANG:95,LEE:96,SEIJO:92,ORDEJON:95,STEWART:96,SHUKLA:98,HELGAKER:00,HEAD-GORDON:03}
which is a common step to first-principles and semiempirical calculations.
In low order scaling energy minimization methods,
the traditional cubic scaling diagonalization of the
matrix representation of the
effective one-electron Hamiltonian in a finite basis set,~\cite{ROOTHAAN:51}
which leads to the canonical orbitals,
is substituted by different algorithms which,
either solve directly for the unique optimal density
matrix,~\cite{YANG:91,YANG:95,LEE:96,HELGAKER:00,HEAD-GORDON:03}
or for some sort of arbitrary optimal localized
orbitals.~\cite{ORDEJON:96,SEIJO:92,STEWART:96,SHUKLA:98}
In parallel to linear scaling electronic structure methods, a
significant development has also been made in embedding methods,
which focus the computational effort on local properties of a 
system~\cite{BARANDIARAN:88,SEIJO:99}
(see Refs.~\onlinecite{Whitten96} and \onlinecite{GOVIND:99} for 
recent reviews).

In this paper,
we present a new method for the energy minimization step
which can be used in semiempirical and first-principles
Hartree-Fock and Kohn-Sham calculations.
It is expected to be useful as a complement of linear-scaling methods 
for the computation of the Hamiltonian matrix,
which are nowadays available for local exchange-correlation potentials
and Coulomb potentials,~\cite{WHITE:94,WHITE:96,ORDEJON:96,STRAIN:96,SCUSERIA:99}
as well as for exact exchange fields.~\cite{OCHSENFELD:98}
The energy minimization method is based on exploiting
the self-consistent calculation of the optimum
localized orbitals of any localization method of choice
and the use of orbital-specific basis sets.
The $n$ occupied localized orbitals of a very large molecule
which correspond to a chosen localization method
(e.g. the popular methods of Boys,~\cite{BOYS:60}
Edmiston-Ruedenberg,~\cite{EDMISTON:63}
and Pipek-Mezey,~\cite{PIPEK:89}
or any other available or newly developed localization method)
can be regarded as a mosaic of orbitals, 
where each of its component \tesserae\ contains only a few orbitals.
In the present method we define an orbital \tessera\ as a 
subset of the occupied localized orbitals (of the method of choice)
which are localized in some region of real space
and compute the localized orbitals of each
\tessera\ out of one specific pseudoeigenvalue equation
to be solved in a basis set expansion approximation,
using a basis set specific to the \tessera,
or orbital-specific basis set.
In other words, a set of building-block embedded cluster
pseudoeigenvalue coupled equations, one for each \tessera, is solved in a
self-consistent manner. Doing so, the method becomes parallel (the
\tesserae\ are computed independently of each other in the
self-consistent procedure) and exhibits a linear-scaling dependence
with the size of the molecule. We call the present method \mosaico.

Early works on localized orbitals proposed the ideas of computing them directly 
by a self-consistent procedure~\cite{EDMISTON:63,GILBERT:64}
and computing one or several localized orbitals 
out of separate eigenvalue equations
starting with a set of 
qualitatively localized orbitals.~\cite{PETERS:69}
These ideas have been used by several authors 
to propose methods leading to some particular sets of localized orbitals
or to localized orbitals dependent on the initial 
guess.~\cite{SEIJO:92,WILHITE:73,KUNZ:73,SCHLOSSER:73,MATSUOKA:77}
We apply them to the computation of the occupied
localized orbitals of any localization method of choice.
Here, we do not pay attention to the computation of virtual localized orbitals;
methods for the determination of virtual localized orbitals useful
in wave function based correlation methods have been used already in 
early works~\cite{WILHITE:73}.
The idea of using different basis sets for different regions of the system
is also present in early work,~\cite{PETERS:69}
it has been used and discussed by several 
authors,~\cite{ORDEJON:96,SEIJO:92,MATSUOKA:77,STOLL:80}
and is a common procedure in embedded cluster and 
effective core potential calculations.~\cite{SEIJO:99,Whitten96}

Besides its use as a linear scaling method in large molecules and
solids, the \mosaico\ method can be used as an embedded cluster
method, where only a few localized orbitals of a large system are
treated variationally while the rest is taken from a calculation on
a similar molecule and frozen, with obvious computational
advantages. This can be done because the solutions of the
building-block embedded cluster coupled equations are the localized
orbitals corresponding to a given localization method of choice,
which enables their transferability between similar molecules.


In Sec.~\ref{SEC:Method} we present the details of the \mosaico\ method.
We performed monitoring calculations
on large poly(ethylen oxide) molecules and 
three dimensional carbon monoxide clusters
using an extended H\"{u}ckel semiempirical Hamiltonian,
which are presented in Sec.~\ref{SEC:Results}.
They are aimed at showing
the convergence of the parallel calculation to the right solution,
the convergence of the total energy with the size of the orbital-specific
basis sets towards the exact value,
the linear-scaling characteristics of the method, and
the performance of embedded cluster approximations.

\section{METHOD}
\label{SEC:Method}

\subsection{Basics of the method}

Let us consider
the Hartree-Fock equations in wave function based {\it ab initio}
methods,~\cite{ROOTHAAN:51,FOCK:30,HARTREE:35}
the Kohn-Sham equations in density functional theory,~\cite{KOHN:65}
or the effective Hartree-Fock equations in semiempirical
methods.~\cite{HOFFMANN:63}
The canonical form of the spin-restricted closed-shell version
of these equations can be written as
\begin{eqnarray}
  \label{EQ:caneq}
  \hat{F} \underline{\varphi}^{can}
    =
  \underline{\varphi}^{can} \,\underline{\varepsilon}
\,,
\end{eqnarray}
with an appropriate choice of the one-electron Hamiltonian $\hat{F}$
for each case,
where $\underline{\varphi}^{can}$ is a row vector of $n$ occupied molecular orbitals,
\begin{eqnarray}
  \label{EQ:canmos}
  \underline{\varphi}^{can}
    =
  \left( \mid \varphi^{can}_1 \rangle,
         \mid \varphi^{can}_2 \rangle,
         \ldots,
         \mid \varphi^{can}_n \rangle \right)
\,,
\end{eqnarray}
and
$\underline{\varepsilon}$ is an
$n \times n$ diagonal matrix of orbital energies.
This and what follows
can be generalized to the spin-unrestricted cases if the orbitals
and the Hamiltonian in Eqs.~\ref{EQ:caneq} and \ref{EQ:canmos} are adequately
substituted by the $\alpha$ and $\beta$ choices;~\cite{POPLE:54}
here we will continue with the detailed description of the
spin-restricted closed-shell case for the sake of clarity.
The virtual orbitals, $\underline{\varphi}^{vir}$,
are also solutions of Eq.~\ref{EQ:caneq};
in this paper, however,
we will focus our attention on the occupied spectrum and,
unless specified,
we will refer only to different sets of occupied orbitals from now on.

The Fock-Dirac one-electron density operator is defined as
\begin{eqnarray}
\label{EQ:rho}
  \hat{\rho}
    =
  \underline{\varphi}^{can} \underline{\varphi}^{can \dagger}
    =
  \sum_{i=1}^{n} \mid \varphi^{can}_i \rangle \langle \varphi^{can}_i  \mid
\,.
\end{eqnarray}
It is invariant under
arbitrary unitary transformations of the occupied orbitals,
\begin{eqnarray}
\label{EQ:canUL}
  \underline{\varphi}^{L}
    =
  \underline{\varphi}^{can} \,_ {can}\underline{U}^{L}
\,, \\
  \hat{\rho}
    =
  \underline{\varphi}^{can} \underline{\varphi}^{can \dagger}
    =
  \underline{\varphi}^{L} \underline{\varphi}^{L \dagger}
\,,
\end{eqnarray}
where we use the notation
$_{can}\underline{U}^{L}$ for the unitary matrix that transforms
canonical occupied 
orbitals onto localized orbitals of a given localization method,
which we label $L$.
Also, $\hat{\rho}$ is the projection operator of the occupied space,
\begin{eqnarray}
\label{EQ:rhoisproj}
  \hat{\rho}\,\underline{\varphi}^{can} = \underline{\varphi}^{can}
\,, \nonumber \\
  \hat{\rho}\,\underline{\varphi}^{L} = \underline{\varphi}^{L}
\,, \\ \nonumber
  \hat{\rho}\,\underline{\varphi}^{vir} = \underline{0}
\,.
\end{eqnarray}

The fact that wave function, one-electron density, total energy,
and Fock operator are invariant under
unitary transformations within the occupied space (Eq.~\ref{EQ:canUL})
has been exploited to define localized orbitals,
which are useful to facilitate large scale calculations
and to bridge the gap between
extensive numerical calculations and qualitative chemical thinking.
A localization method, say $L$, can be defined by its choice of
$_{can}\underline{U}^{L}$.
Very sound and popular localization methods
are the methods of Boys,~\cite{BOYS:60}
Edmiston-Ruedenberg,~\cite{EDMISTON:63}
and Pipek-Mezey,~\cite{PIPEK:89}
although others have been proposed.
All of the above can be used in first-principles methods;
Pipek-Mezey's can also be applied in semiempirical methods.
The common procedure to compute localized orbitals
is to complete firstly a canonical calculation and,
later, use the canonical orbitals in an iterative optimization process
converging to $_{can}\underline{U}^{L}$.
Gilbert~\cite{GILBERT:64} has pointed out that
any set of occupied localized orbitals formally fulfills the eigenvalue equation
of an effective Fock (or Kohn-Sham) Hamiltonian defined as
$\hat{F}^{L} = \hat{F} - \hat{\rho}\,\hat{F}\,\hat{\rho}
+ \hat{\rho}\,\hat{L}\,\hat{\rho}$,
\begin{eqnarray}
\label{EQ:loceq}
  \hat{F}^{L}\,\underline{\varphi}^{L}
    =
  \left[ \hat{F} - \hat{\rho}\,\hat{F}\,\hat{\rho}
                 + \hat{\rho}\,\hat{L}\,\hat{\rho}
  \right] \,\underline{\varphi}^{L}
    =
  \underline{\varphi}^{L} \, \underline{\lambda}
\,,
\end{eqnarray}
where $\hat{L}$ is a Hermitean localization operator
and $\underline{\lambda}$ is a diagonal matrix whose diagonal elements are
the eigenvalues of $\hat{F}^{L}$ and of $\hat{L}$,
\begin{eqnarray}
  \underline{\lambda} =
      \underline{\varphi}^{L \dagger}
      \,\hat{L}\,
      \underline{\varphi}^{L}
\,.
\end{eqnarray}
Localization operators $\hat{L}$ corresponding to the above mentioned
localization methods are, in general, not known
and Eq.~\ref{EQ:loceq} has not been exploited
to compute the localized orbitals via diagonalization of the matrix
$\underline{\varphi}^{can \dagger} \,\hat{L}\, \underline{\varphi}^{can}$,
to our knowledge.

Eq.~\ref{EQ:loceq} has been discussed by several
authors~\cite{SEIJO:92,EDMISTON:63,MATSUOKA:77}
and it has been used as a basis for
a building-block technique based on a self-consistent series of
embedded cluster calculations.~\cite{SEIJO:92}
The iterative solution of
the building-block equations of Ref.~\onlinecite{SEIJO:92},
which contain arbitrary localization operators,
leads to localized orbitals dependent on the initial guess and
on the iteration procedure.
Although in principle this is not detrimental for total energy calculations,
it is an undesirable property because it creates problems of transferability
of localized orbitals between similar molecules.
 
Here, on the basis of the same ideas than in Ref.~\onlinecite{SEIJO:92},
we present a new building-block and embedding method
that, starting from an arbitrary guess
and a choice of a particular localization method
(Boys', Edmiston-Ruedenberg's, Pipek-Mezey's, or any other),
leads,
in a controlled manner,
to the corresponding localized orbitals.
We will call this method \mosaico.


Let us suppose we set the goal of computing the $n$ occupied localized orbitals
corresponding to a localization method $L$ for the ground state of a molecule,
\begin{eqnarray}
\label{EQ:lmos}
  \underline{\varphi}^{L}
    =
  \left( \mid \varphi^{L}_1 \rangle,
         \mid \varphi^{L}_2 \rangle,
         \ldots,
         \mid \varphi^{L}_n \rangle
  \right)
\,,
\end{eqnarray}
and related properties such as electron density and total energy.
We see the whole set of localized orbitals as a mosaic of orbitals,
and define subsystem, fragment, cluster, or {\it tessera} as a
subset of these orbitals which are localized in some region of real
space. (For example, in an organic acid R-COOH we may define one of
the subsystems or {\it tesserae} as that made of nine orbitals
localized in the spatial region close to the COOH nuclei.) The terms
subsystem, fragment, and cluster have been used by many authors with
different meanings and we prefer to use the term {\it tessera} all
over the paper for the present definition of subsystem. 
In this way,
the whole mosaic of localized orbitals is made of
$N$ {\it tesserae} \mbox{$A$, $B$, \ldots},
and the vector of localized orbitals can be rewritten as
\begin{eqnarray}
\label{EQ:phiL}
  \underline{\varphi}^{L}
    =
  \left(
     \underline{\varphi}^{L}_{A},
     \underline{\varphi}^{L}_{B},
     \ldots
  \right)
\,,
\end{eqnarray}
where
$\underline{\varphi}^{L}_{A}$
is a row vector with the $n_A$ occupied orbitals of \tessera\ A,
\begin{eqnarray}
  \underline{\varphi}^{L}_{A}
  =
  \left( \mid \varphi^{L}_{A1} \rangle,
         \mid \varphi^{L}_{A2} \rangle,
         \ldots,
         \mid \varphi^{L}_{An_A} \rangle \right)
\,,
\end{eqnarray}
$\underline{\varphi}^{L}_{B}$
a row vector with the $n_B$ orbitals of \tessera\ B,
\begin{eqnarray}
  \underline{\varphi}^{L}_{B}
  =
  \left( \mid \varphi^{L}_{B1} \rangle,
         \mid \varphi^{L}_{B2} \rangle,
         \ldots,
         \mid \varphi^{L}_{Bn_B} \rangle \right)
\,,
\end{eqnarray}
and so on,
with $n_A + n_B + \ldots = n$.
The orbitals of a \tessera\ define a subspace of the occupied space
whose density and projection operator is
\begin{eqnarray}
  \label{EQ:rhoA}
  \hat{\rho}_A =
  \underline{\varphi}^{L}_A \,\underline{\varphi}^{L \dagger}_A
\,.
\end{eqnarray}
The density operators of the \tesserae\ fulfil
\begin{eqnarray}
\label{EQ:rhocond-a}
  \hat{\rho}_A \hat{\rho}_B = \delta_{AB} \hat{\rho}_A
\,, \\
\label{EQ:rhocond-b}
  \hat{\rho} = \sum_{B=1}^{N} \hat{\rho}_B
\,, \\
\label{EQ:rhocond-c}
  \hat{\rho} \hat{\rho}_A = \hat{\rho}_A \hat{\rho} = \hat{\rho}_A
\,,
\end{eqnarray}
where the sum in Eq.~\ref{EQ:rhocond-b} extends over all the $N$
\tesserae\ of the system.

In the \mosaico\ method
we seek to compute the localized orbitals of each \tessera\
out of its own eigenvalue equation:
\begin{eqnarray}
  \label{EQ:bb-a}
  \mbox{\it tessera}\,\,A &:&
     \nonumber \\ 
  \hat{F}^{L}_A\,\underline{\varphi}^{L}_A
    &=&
    \left[ \hat{F} - \hat{\rho}\,\hat{F}\,\hat{\rho}
                        + \hat{\rho}\,\hat{L}_A\,\hat{\rho} \right]
       \,\underline{\varphi}^{L}_A
    =
    \underline{\varphi}^{L}_A\,\underline{\lambda}_A
\,,           \\
  \label{EQ:bb-b}
  \mbox{\it tessera}\,\,B &:&
     \nonumber \\ 
  \hat{F}^{L}_B\,\underline{\varphi}^{L}_B
    &=&
    \left[ \hat{F} - \hat{\rho}\,\hat{F}\,\hat{\rho}
                        + \hat{\rho}\,\hat{L}_B\,\hat{\rho} \right]
       \,\underline{\varphi}^{L}_B
    =
    \underline{\varphi}^{L}_B\,\underline{\lambda}_B
\,, \\ \nonumber
  & \ldots & \hfill \hspace{1mm}
\end{eqnarray}
($\underline{\lambda}_A$, $\underline{\lambda}_B$, \ldots, being
diagonal matrices of size $n_A \times n_A$, $n_B \times n_B$,
\ldots,) under the conditions of Eqs.~\ref{EQ:rhoA}-\ref{EQ:rhocond-c}.
These conditions are fulfilled if all the orbitals are
eigenfunctions of the same Hermitean operator. In other words, all
the orbitals $\underline{\varphi}^{L}_B$ for $B \ne A$ must be
eigenfunctions of $\hat{F}^{L}_A$, all the orbitals
$\underline{\varphi}^{L}_A$ for $A \ne B$ must be eigenfunctions of
$\hat{F}^{L}_B$, and so on. This means that the subsystem
localization operators $\hat{L}_A$, $\hat{L}_B$, \ldots\ of
Eqs.~\ref{EQ:bb-a}, \ref{EQ:bb-b}, \ldots\ must be such that the
effective subsystem Hartree-Fock or Kohn-Sham Hamiltonians
$\hat{F}^{L}_A$, $\hat{F}^{L}_B$, \ldots, have the same
eigenfunctions as $\hat{F}^{L}$ in Eq.~\ref{EQ:loceq} but different
eigenvalues.
(Note that the virtual orbitals $\underline{\varphi}^{vir}$
are also eigenfunctions of Eqs.~\ref{EQ:bb-a}, \ref{EQ:bb-b}, \ldots\
since $\hat{\rho} \underline{\varphi}^{vir} = \underline{0}$;
their calculation has not been stated as a goal here and they will
not be referred to in the rest of the Section.)

As commented above,
the localization operator $\hat{L}$ in Eq.~\ref{EQ:loceq}
corresponding to a localization method $L$
is usually not known.
However,
every localization method $L$ has a well defined procedure
for the computation of the unitary transformation matrix
$_{can}\underline{U}^{L}$ (Eq.~\ref{EQ:canUL})
in a given molecule.~\cite{BOYS:60,EDMISTON:63,PIPEK:89}
This procedure can also be applied
to any orthogonal basis of the occupied space $\underline{\varphi}^{(0)}$
other than the canonical orbital basis $\underline{\varphi}^{can}$;
the result is then the unitary matrix $_{0}\underline{U}^{L}$
that transforms the initial non-canonical set of occupied orbitals
onto the $L$-method localized orbitals,
\begin{eqnarray}
  \label{EQ:0UL}
  \underline{\varphi}^{L}
    =
    \underline{\varphi}^{(0)}\,\,_{0}\underline{U}^{L}
\,.
\end{eqnarray}
At this point, we can express the operator $\hat{L}$ of Eq.~\ref{EQ:loceq}
as its spectral representation in any basis of the occupied space,
e.g. $\underline{\varphi}^{(0)}$,
\begin{eqnarray}
  \hat{L}
          = \underline{\varphi}^{L}
            \,\underline{\lambda}\,
            \underline{\varphi}^{L \dagger}
          = \underline{\varphi}^{(0)}
            \,_{0}\underline{U}^{L}\,
            \,\underline{\lambda}\,
            \,\,_{0}\underline{U}^{L \dagger}\,
            \underline{\varphi}^{(0) \dagger}
\,.
\end{eqnarray}

Now, consistently with the discussion following
Eqs.~\ref{EQ:bb-a}, \ref{EQ:bb-b}, \ldots,
we can define the following Hermitean
subsystem or \tessera\ localization operators:
\begin{eqnarray}
\label{EQ:la}
  \hat{L}_A & = &
            \underline{\varphi}^{L}
            \,\underline{\lambda}_{A(n)}\,
            \underline{\varphi}^{L \dagger}
          =
            \underline{\varphi}^{(0)}
            \,_{0}\underline{U}^{L}\,
            \,\underline{\lambda}_{A(n)}\,
            \,\,_{0}\underline{U}^{L \dagger}\,
            \underline{\varphi}^{(0) \dagger}
\,, \\
\label{EQ:lb}
  \hat{L}_B & = &
            \underline{\varphi}^{L}
            \,\underline{\lambda}_{B(n)}\,
            \underline{\varphi}^{L \dagger}
          =
            \underline{\varphi}^{(0)}
            \,_{0}\underline{U}^{L}\,
            \,\underline{\lambda}_{B(n)}\,
            \,\,_{0}\underline{U}^{L \dagger}\,
            \underline{\varphi}^{(0) \dagger}
\,, \\ \nonumber
   & \ldots &
\,
\end{eqnarray}
where $\underline{\lambda}_{A(n)}$, $\underline{\lambda}_{B(n)}$,
\ldots\ are diagonal matrices of size $n \times n$ with arbitrary
diagonal real data. A choice consisting on low values for the $n_A$
diagonal elements of $\underline{\lambda}_{A(n)}$ corresponding to
the localized orbitals of \tessera\ $A$,
$\underline{\varphi}^{L}_A$, and sufficiently higher values for the
$n - n_A$ remaining diagonal elements, guarantees that the localized
orbitals of \tessera\ $A$ are computed as the lowest $n_A$
eigenvectors of the effective Hartree-Fock or Kohn-Sham Hamiltonian
$\hat{F}^{L}_A$  (Eq.~\ref{EQ:bb-a}), which is a convenient choice
for safe and efficient orbital selection in the iterations of
self-consistent procedures. Obviously, the same comments stand for
all \tesserae. We may remark that a choice consisting on negative
values for the $n_A$, $n_B$, \ldots\ cited diagonal elements and
zero values for the $n-n_A$, $n-n_B$, \ldots\ remaining elements,
although not necessary, 
is efficient and simplifies the expression of the subsystem
localization operators,
\begin{eqnarray}
  \label{EQ:la2}
  \hat{L}_A & = &
            \underline{\varphi}^{L}_{A}
            \,\underline{\lambda}_{A}\,
            \underline{\varphi}^{L \dagger}_{A}
          =
            \underline{\varphi}^{(0)}
            \,_{0}\underline{U}^{L(A)}\,
            \,\underline{\lambda}_{A}\,
            \,\,_{0}\underline{U}^{L(A) \dagger}\,
            \underline{\varphi}^{(0) \dagger}
\,, \\
  \hat{L}_B & = &
            \underline{\varphi}^{L}_{B}
            \,\underline{\lambda}_{B}\,
            \underline{\varphi}^{L \dagger}_{B}
          =
            \underline{\varphi}^{(0)}
            \,_{0}\underline{U}^{L(B)}\,
            \,\underline{\lambda}_{B}\,
            \,\,_{0}\underline{U}^{L(B) \dagger}\,
            \underline{\varphi}^{(0) \dagger}
\,, \\ \nonumber
   & \ldots &
\,
\end{eqnarray}
where
$\underline{\lambda}_{A}$, of size $n_A \times n_A$,
is the diagonal eigenvalue matrix of Eq.~\ref{EQ:bb-a},
and
$_{0}\underline{U}^{L(A)}$ is a rectangular matrix made of
$n_A$ columns of $_{0}\underline{U}^{L}$,
and similarly for all \tesserae.

Using Eqs.~\ref{EQ:la}, \ref{EQ:lb}, \ldots\
in \ref{EQ:bb-a}, \ref{EQ:bb-b}, \ldots,
plus the fact that $\hat{\rho}$ is the projection operator
of the occupied space (Eq.~\ref{EQ:rhoisproj}),
results in the working equations of the \mosaico\ method
for a localization method of choice $L$:
\begin{eqnarray}
\label{EQ:mosaico-a}
  \mbox{\it tessera}\,\,A &:& \nonumber \\ 
  \hat{F}^{L}_A\,\underline{\varphi}^{L}_A
    &=&
    \left[ \hat{F} - \hat{\rho}\,\hat{F}\,\hat{\rho} +
            \underline{\varphi}^{(0)}
            \,_{0}\underline{U}^{L}\,
            \,\underline{\lambda}_{A(n)}\,
            \,\,_{0}\underline{U}^{L \dagger}\,
            \underline{\varphi}^{(0) \dagger}
    \right]
       \,\underline{\varphi}^{L}_A
                              \nonumber \\
    &=&
    \underline{\varphi}^{L}_A\,\underline{\lambda}_A
\,,           \\
\label{EQ:mosaico-b}
  \mbox{\it tessera}\,\,B &:& \nonumber \\ 
  \hat{F}^{L}_B\,\underline{\varphi}^{L}_B
    &=&
    \left[ \hat{F} - \hat{\rho}\,\hat{F}\,\hat{\rho} +
            \underline{\varphi}^{(0)}
            \,_{0}\underline{U}^{L}\,
            \,\underline{\lambda}_{B(n)}\,
            \,\,_{0}\underline{U}^{L \dagger}\,
            \underline{\varphi}^{(0) \dagger}
    \right]
       \,\underline{\varphi}^{L}_B
                              \nonumber \\
    &=&
    \underline{\varphi}^{L}_B\,\underline{\lambda}_B
\,, \\ \nonumber
  & \ldots & 
\end{eqnarray}
Eq.~\ref{EQ:mosaico-a} is the Mosaico equation for the embedded
\tessera\ $A$,
Eq.~\ref{EQ:mosaico-b} for the embedded \tessera\ $B$,
and so on.
They are pseudoeigenvalue equations
(even in the case of semiempirical methods with
density-independent $\hat{F}$ Hamiltonians.)
They can be solved with standard SCF iterative procedures.
Starting with an initial guess, $\underline{\varphi}^{(0)}$,
the procedure of the localization method of choice, $L$, is applied
in order to compute $_{0}\underline{U}^{L}$,
which, together with $\hat{\rho}$ and $\hat{F}$,
give the embedded \tessera\ Hamiltonians
$\hat{F}^{L}_A$, $\hat{F}^{L}_B$, \ldots\
for the current iteration.
(Note that the elements of the diagonal matrices
$\underline{\lambda}_{A(n)}$, $\underline{\lambda}_{B(n)}$, \ldots,
are input real numbers.)
Solving the eigenvalue equations leads to new orbitals
which are used to update $\underline{\varphi}^{(0)}$ and iterate.
At convergence,
$\underline{\varphi}^{(0)} = \underline{\varphi}^{L}$
and
$_{0}\underline{U}^{L}$ is the unit matrix.
Also, the eigenvalues of the \tesserae\ are identical to the
input values of
$\underline{\lambda}_{A(n)}$, $\underline{\lambda}_{B(n)}$, \ldots.
For instance, the $n_A$ non-zero values of $\underline{\lambda}_{A}$
coincide with the non-zero values of $\underline{\lambda}_{A(n)}$
associated with the localized orbitals of \tessera\ A.

Several options are open for iterative procedures leading to the solutions
of the Mosaico equations \ref{EQ:mosaico-a}, \ref{EQ:mosaico-b}, \ldots\
They all have to face two basic types of iterations:
(1) The microiterations, or \tessera\ iterations,
which are standard SCF iterations addressed to solve one
of the embedded \tessera\ equations, e.g. Eq.~\ref{EQ:mosaico-a},
for fixed orbitals of the other \tesserae.
All the methods available for speeding the solution of the canonical
Hartree-Fock or Kohn-Sham equations can be used here.
(2) The macroiterations, or mosaic iterations,
which involve repeated solutions of all the embedded \tessera\ equations.
The macroiterations can be performed sequentially on a given list
of \tesserae, e.g. $A$, $B$, \ldots, $A$, $B$, \ldots,
meaning that the orbitals computed for \tessera\ $A$ are used
in the calculation of \tessera\ $B$, and so on.
Most interestingly, they can be performed in parallel, 
meaning that the calculations of
all \tesserae\ $A$, $B$, \ldots\ are done using the whole mosaic
of orbitals from the previous macroiteration.
This alternative is very
important, because it allows to take full advantage of parallelism
at a high level of the calculation. In other words, the full \mosaico\
calculation is made of a sequence of macroiterations, each of them
consisting of a parallel set of Hartree-Fock, Kohn-Sham, or
semiempirical calculations on the embedded \tesserae\ $A$, $B$,
\ldots. Each of these \tessera\ calculations, which can be time
consuming, can be performed in a separate processor or computer.

Alternataively,
the macroiterations can be done {\it \`{a} la carte}, that is
restricted to only one or several selected \tesserae.
This is to say that only some of the localized orbitals are optimized
whereas the rest of them are frozen.
This is the embedded cluster approximation.
Its reliability rests upon the transferability of the localized orbitals.
Embedded cluster calculations
are significantly cheaper than the full calculation of a complete system.
They can be performed on a molecule or solid
if a calculation on a similar molecule or solid has been carried out
before in order to provide the orbitals of the embedding frozen \tesserae.
They are specially useful to study defects in solids,
chemisorption,
series of molecules with different substituents,
or reactions taking place in local regions of large molecules.

The \mosaico\ equations
(Eqs.~\ref{EQ:mosaico-a}, \ref{EQ:mosaico-b}, \ldots ),
which give the localized orbitals of a given localization method $L$
and, in consequence, the same total energy and electron density than
the canonical calculation,
are the basis for approximations that make them useful in practice.
These approximations,
which resort to truncations supported by the localized nature of the orbitals,
are systematic and converge to the exact result.
They are basically twofold: 
(1)~An orbital-specific
basis set approximation can be adopted, where the localized orbitals
of a \tessera\ are expanded in a different basis set than the
localized orbitals of another \tessera. This can be done because the
$n_A$ occupied localized orbitals of \tessera\ $A$ are the only
orbitals to be computed by solving equation \ref{EQ:mosaico-a},
whereas the $n-n_A$ orbitals of the other \tesserae\ are discarded;
consequently, a local basis set can be used as long as it is
sufficient for a good representation of the localized orbitals
$\underline{\varphi}^{L}_A$. Obviously, this is true for any
\tessera. 
(2)~A localization algorithm based on local rotations can be used
in each interation instead of the standard algorithm of
the localization method of choice, $L$. 
In the local rotations
algorithm, the localized orbitals of \tessera\ $A$ are computed in
each iteration using a subset of the $\underline{\varphi}^{(0)}$
orbitals which is localized in \tesserae\ not too distant from $A$.
This can be done because the target localized orbitals are computed
out of input localized orbitals rather than out of canonical
orbitals.

\subsection{Use of orbital-specific basis sets, OSBS}
\label{SEC:OSBS}

In a standard molecular calculation, all the orbitals of a molecule are
expanded in a common basis set,
which is the basis set of the molecule
and consists of $n_{BSF}$ basis set functions,
\begin{eqnarray}
  \underline{\chi}
    =
  \left( \mid \chi_{1} \rangle,
         \mid \chi_{2} \rangle,
         \ldots,
         \mid \chi_{n_{BSF}} \rangle \right)
\,,
\end{eqnarray}
which could be contracted Gaussian functions or other sort of local
functions. In the orbital-specific basis set approximation, OSBS,
the localized orbitals of \tessera\ $A$ are expanded in a local
basis set made of $b_A$ functions, the localized orbitals of
\tessera\ $B$ in a local basis set made of $b_B$ functions, and so
on. Although not necessary, it is very convenient that the local
basis sets are subsets of the global basis set, and that several
\tesserae\ share a number of basis set functions. The local basis
sets can be represented by the following row vectors,
\begin{eqnarray}
  \label{EQ:chia}
  \mbox{\it tessera}\,\,A &:& \nonumber \\ 
  \underline{\chi}_A
    &=&
  \left( \mid \chi_{A1} \rangle,
         \mid \chi_{A2} \rangle,
         \ldots,
         \mid \chi_{Ab_A} \rangle \right)
\,, \\
  \label{EQ:chib}
  \mbox{\it tessera}\,\,B &:& \nonumber \\ 
  \underline{\chi}_B
    &=&
  \left( \mid \chi_{B1} \rangle,
         \mid \chi_{B2} \rangle,
         \ldots,
         \mid \chi_{Bb_B} \rangle \right)
\,, \\ \nonumber
  & \ldots &
\end{eqnarray}
Accordingly, eqs.~\ref{EQ:mosaico-a}, ~\ref{EQ:mosaico-b}, \ldots,
take the usual matrix form,~\cite{ROOTHAAN:51}
\begin{eqnarray}
  \label{EQ:a}
  \mbox{\it tessera}\,\,A &:& 
  \underline{F}^{L}_{A}\,\underline{C}^{L}_A
    =
      \underline{S}_{A}\,\underline{C}^{L}_A\,\underline{\lambda}_A
\, \\
  \label{EQ:b}
  \mbox{\it tessera}\,\,B &:& 
  \underline{F}^{L}_{B}\,\underline{C}^{L}_B
    =
      \underline{S}_{B}\,\underline{C}^{L}_B\,\underline{\lambda}_B
\,, \\ \nonumber
  & \ldots &
\end{eqnarray}
In Eq.~\ref{EQ:a}, for instance,
$\underline{F}^{L}_{A}$ and $\underline{S}_{A}$ are the
$b_A \times b_A$
matrix representations of the $\hat{F}^{L}_{A}$ and the unit operators
in the $\underline{\chi}_A$ basis set,
\begin{eqnarray}
  \underline{F}^{L}_{A}
    =
      \underline{\chi}_A^{\dagger}\,\hat{F}^{L}_{A}\,\underline{\chi}_A
\,,\hspace{10mm}\,
  \underline{S}^{L}_{A}
    =
      \underline{\chi}_A^{\dagger}\,\underline{\chi}_A
\,,
\end{eqnarray}
and $\underline{C}^{L}_A$ is the $b_A \times n_A$ matrix of
localized orbital coefficients,
\begin{eqnarray}
\label{EQ:neworb}
  \underline{\varphi}^{L}_{A}
    =
      \underline{\chi}_A\,\underline{C}^{L}_A
\,.
\end{eqnarray}
The solution of Eq.~\ref{EQ:a} can be attained using
standard \order($b_A^3$) diagonalization procedures.
Except for low gap materials,
where the degree of localization attainable is limited
and the localized orbitals may decay slowly with
distance,~\cite{KOHN:73}
the local basis set size $b_A$ is expected to remain within reasonable limits
for a diagonalization or, more generally, for an embedded cluster calculation.
Obviously, the growth of $b_A$ will impose more practical limitations
in 3D systems, like bulk solids and very big clusters,
than in 2D and 1D systems, like many molecules.
In general, all the methods useful to speed up a standard molecular
calculation, like convergence acceleration methods, can be used with
Eq.~\ref{EQ:a}. But, in this case, additional advantage can be taken
from the fact that only a small number of eigenvalues/eigenfunctions
are computed for each \tessera, and efficient diagonalization
algorithms used in CI calculations, like the multiroot Davidson-Liu
method,~\cite{DAVIDSON:75,LIU:78} can be applied to significantly
reduce the prefactor of the \order($b_A^3$) dependence.

Let us comment on the calculation of the effective Hamiltonian
matrix of a \tessera, $\underline{F}^{L}_{A}$, where the localized
nature of the orbitals is also profitable. 
For simplicity, we
will call $\underline{\phi}$ the row vector of the $n$ current
localized orbitals ($\underline{\varphi}^{L}$, Eq.~\ref{EQ:0UL}) 
at a given iteration,
which can be written as the union of the row vectors
$\underline{\phi}_A$, $\underline{\phi}_B$, \ldots,
that contain the $n_A$, $n_B$, \ldots\
current localized orbitals of
\tessera\ $A$, $B$, \ldots\ respectively,
\begin{eqnarray}
  \underline{\phi} = \left(
      \underline{\phi}_A, \underline{\phi}_B, \ldots \right)
\,.
\end{eqnarray}
The $b_A \times b_A$
effective Hamiltonian matrix of \tessera\ $A$ is
\begin{eqnarray}
\label{EQ:FeffA}
  \underline{F}^{L}_{A}
    &=&
      \underline{\chi}_A^{\dagger}\,\hat{F}^{L}_{A}\,\underline{\chi}_A
    \nonumber \\
    &=&
      \underline{\chi}_A^{\dagger}\,\hat{F}\,\underline{\chi}_A
      -
      \underline{\chi}_A^{\dagger}\,
        \hat{\rho}\hat{F}\hat{\rho}
      \,\underline{\chi}_A
      +
      \underline{\chi}_A^{\dagger}\,
            \underline{\phi}
            \,\underline{\lambda}_{A(n)}\,
            \underline{\phi}^{\dagger}
      \,\underline{\chi}_A
\,.
\end{eqnarray}
The first term in the right hand side of Eq.~\ref{EQ:FeffA}
is a diagonal block of the Hamiltonian matrix of the full system.
The second term can be expanded in inter-\tessera\ terms
by inserting Eq.~\ref{EQ:rhocond-b}:
\begin{eqnarray}
\label{EQ:rhohrho}
&&  \underline{\chi}_A^{\dagger}
    \,\hat{\rho}\hat{F}\hat{\rho}\,
  \underline{\chi}_A
    =   \nonumber \\
&&  \sum_{B=1}^{N} \sum_{C=1}^{N} 
    \left( \underline{\chi}_A^{\dagger}\,\underline{\phi}_B \right)
    \left( \underline{\phi}_B^{\dagger}\,\hat{F}\,\underline{\phi}_C \right)
    \left( \underline{\phi}_C^{\dagger}\,\underline{\chi}_A \right)
\,.
\end{eqnarray}
In the evaluation of Eq.~\ref{EQ:rhohrho},
we can take advantage of the locality of basis sets and molecular orbitals
by evaluating a
$\Delta^S_{AB}$
interaction table and a
$\Delta^F_{AB}$
interaction table.
$\Delta^S_{AB}$ is set to $0$
if all the elements of the $b_A \times n_B$ matrix
$\underline{\chi}_A^{\dagger}\,\underline{\phi}_B$
and all the elements of the $b_B \times n_A$ matrix
$\underline{\chi}_B^{\dagger}\,\underline{\phi}_A$
have an absolute value lower than a given threshold,
and is set to $1$ otherwise.
Similarly,
$\Delta^F_{AB}$ is set to $0$
if all the elements of the $n_A \times n_B$ matrix
$\underline{\phi}_A^{\dagger}\,\hat{F}\,\underline{\phi}_B$
have an absolute value lower than a given threshold,
and is set to $1$ otherwise.
Rearranging Eq.~\ref{EQ:rhohrho} and
using these interaction tables, we can write
\begin{eqnarray}
\label{EQ:rhoHrho-ok}
&&  \underline{\chi}_A^{\dagger}\,
    \hat{\rho}\hat{F}\hat{\rho}
  \,\underline{\chi}_A
   = 
    \sum_{B=1}^{N} \Delta^S_{AB}
    \left( \underline{\chi}_A^{\dagger}\,\underline{\phi}_B \right)
    \left( \underline{\phi}_B^{\dagger}\,\hat{F}\,\underline{\phi}_B \right)
    \left( \underline{\phi}_B^{\dagger}\,\underline{\chi}_A \right)
\nonumber \\
&&  +  
    \sum_{B=2}^{N} \Delta^S_{AB}
    \sum_{C=1}^{B-1} \Delta^S_{AC} \Delta^F_{BC} 
         \Big[
    \left( \underline{\chi}_A^{\dagger}\,\underline{\phi}_B \right)
    \left( \underline{\phi}_B^{\dagger}\,\hat{F}\,\underline{\phi}_C \right)
    \left( \underline{\phi}_C^{\dagger}\,\underline{\chi}_A \right)
\nonumber \\
&&  + \mbox{adjoint}
          \Big]
\,.
\end{eqnarray}
The use of the \tesserae\ interaction tables $\Delta^S_{AB}$ and
$\Delta^F_{AB}$ guarantees that the calculation of the present term
does not scale as $N^2$ because the number of \tesserae\ in the
neighborhood of (or interacting with)
\tessera\ $A$ does not increase indefinitely with
the size of the molecule. Also, note that the interaction tables do
not need to be updated every macroiteration, because the localized
orbitals do not experience big changes in size after a few
iterations.
It should be noticed that
we assume that a linear-scaling method is used for the computation of the
Hamiltonian matrix of the full system,~\cite{WHITE:96,SCUSERIA:99,SOLER:02}
of which diagonal and nondiagonal blocks are needed in 
Eqs.~\ref{EQ:FeffA} and \ref{EQ:rhoHrho-ok}.

The last term in Eq.~\ref{EQ:FeffA} can be written as
\begin{eqnarray}
  \label{EQ:thirdterm}
&&    \underline{\chi}_A^{\dagger}\,
            \underline{\phi}
            \,\underline{\lambda}_{A(n)}\,
            \underline{\phi}^{\dagger}
      \,\underline{\chi}_A
        = 
\nonumber \\ &&
        \sum_{B=1}^{N} \Delta^S_{AB}
        \left( \underline{\chi}_A^{\dagger}\,\underline{\phi}_B \right)
            \,\underline{\lambda}_{A(n)B}\,
        \left( \underline{\phi}_B^{\dagger}\,\underline{\chi}_A \right)
\,,
\end{eqnarray}
where the $\Delta^S_{AB}$ interaction table is used
and $\underline{\lambda}_{A(n)B}$ is the $n_B \times n_B$ diagonal
matrix  resulting from the elements of $\underline{\lambda}_{A(n)}$
corresponding to the localized orbitals of \tessera\ $B$.
For the particular choice of the arbitrary diagonal matrix
$\underline{\lambda}_{A(n)}$ leading to Eq.~\ref{EQ:la2},
Eq.~\ref{EQ:thirdterm} is further simplified:
\begin{eqnarray}
      \underline{\chi}_A^{\dagger}\,
            \underline{\phi}
            \,\underline{\lambda}_{A(n)}\,
            \underline{\phi}^{\dagger}
      \,\underline{\chi}_A
       & = &
        \underline{\chi}_A^{\dagger}\,\underline{\phi}_A
            \,\underline{\lambda}_{A}\,
        \underline{\phi}_A^{\dagger}\,\underline{\chi}_A
\nonumber \\
       & = &
        \underline{\chi}_A^{\dagger}\,\left(
          \sum_{i=1}^{n_A}
            \,\mid {\phi}_A \rangle
            \,{\lambda}_{Ai}\,
            \,\langle {\phi}_A \mid
        \right)\,\underline{\chi}_A
\,,
\end{eqnarray}
where all $\lambda_{Ai}$ must be negative.

\begin{figure}
  \begin{center}
  \mbox{\resizebox{11.6cm}{!}{\includegraphics*{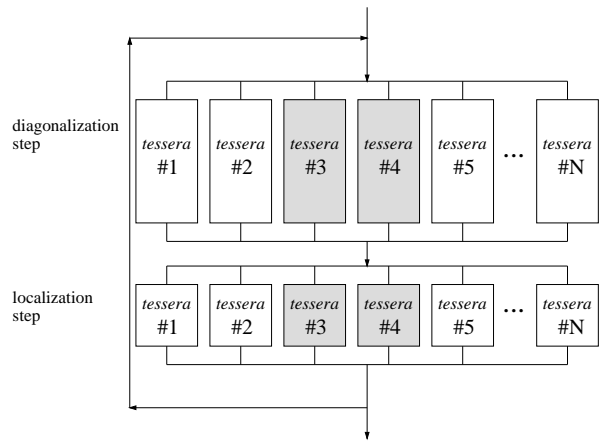}}}
  \end{center}
  \mbox{}\\[-24pt]
  \caption{
           Diagram of the two basic parallel loops of a macroiteration.
           In the first loop,
           the embedded \tessera\ effective Hamiltonian matrices
           are computed and diagonalized
           (Eqs.~\protect\ref{EQ:a}, \mbox{\protect\ref{EQ:b}, \ldots )}.
           In the second loop,
           the $_0\underline{U}^{L}_{CA}$ matrices of the
           localization method of choice, $L$, are computed
           out of the orbitals resulting from the first loop
           and the localization transformations
           (Eqs.~\protect\ref{EQ:a}, \mbox{\protect\ref{EQ:b}, \ldots )}
           are performed.
           The highlighted boxes would be the only to be entered
           in a embedded cluster calculation where \tesserae\
           \#3 and \#4 define the active cluster.
          }
  \label{FIG:LOOPS}
\end{figure}

The computation of the $\underline{F}^{L}_{A}$ matrices
(Eqs.~\ref{EQ:FeffA}, \ref{EQ:rhoHrho-ok}, and \ref{EQ:thirdterm})
and their diagonalization (Eq.~\ref{EQ:a})
can be performed in parallel. This is graphically indicated in
the upper part of Fig.~\ref{FIG:LOOPS}.
The global scaling of this diagonalization step is
$\sum_{B=1}^{N} {\cal O}(b_B^3)$;
in the case of all \tesserae\ having the same local basis set size, $\bar{b}$,
and same number of interactions with other \tesserae,
the scaling is ${\cal O}(\bar{b}^3 N)$.

\subsection{Localization by local rotations}
\label{SEC:LR}

Usually, a given set of localized orbitals of a
molecule, $\underline{\varphi}^{L}$, is computed out of the
canonical orbitals (Eq.~\ref{EQ:canUL}). Well defined algorithms are
routinely used to calculate the $n \times n$ unitary transformation
matrix $_{can}\underline{U}^{L}$, which are normally of order $n^3$
or higher.~\cite{EDMISTON:63,PIPEK:89} In the present method,
however, they are computed in each macroiteration out of other
set of localized orbitals (Eq.~\ref{EQ:0UL}). We can expect the
contributions of the initial localized orbitals
$\underline{\varphi}^{(0)}$ to the target localized orbitals
$\underline{\varphi}^{L}$ to decay with distance. Accordingly, a
$\Delta^{LR}_{AB}$ local rotation table can be computed, where
$\Delta^{LR}_{AB} = 1$ if the initial localized orbitals of
\tessera\ $B$ are used to compute the target localized orbitals of
\tessera\ $A$ (and viceversa) and  $\Delta^{LR}_{AB} = 0$ otherwise.
Several options to compute the $\Delta^{LR}_{AB}$ local rotation
table are possible. Reasonable choices are to use the same criterium
as for the $\Delta^{F}_{AB}$ interaction table except for the use of
a different threshold or, simply, make the $\Delta^{LR}_{AB}$ local
rotation table identical to the $\Delta^{F}_{AB}$ interaction table.

\begin{figure}
  \begin{center}
  \mbox{\resizebox{8.5cm}{!}{\includegraphics*{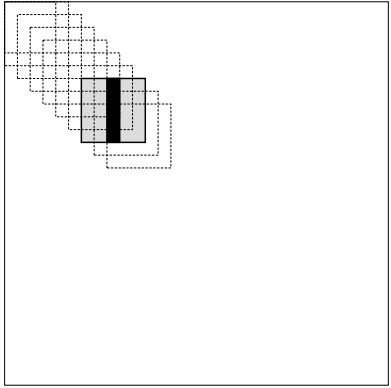}}}
  \end{center}
  \mbox{}\\[-5cm]
  \caption{
           Schematic representation of the localization by local
           rotations (step 2 in Fig.~\protect\ref{FIG:LOOPS}).
           The unitary matrix $_{0}\underline{U}^{L}$ is represented
           together with the blocks that are computed
           in the localization parallel loop
           for \tesserae\ \#1 to \#10.
           The block computed for  \tessera\ \#8
           ($_0\underline{U}^{L}_{LR\#8}$ in Eq.~\protect\ref{EQ:ULRA})
           is highlighted.
           The subblock representing the actual localized orbitals
           of this \tessera\ is represented with a rectangular fill box.
          }
  \label{FIG:LROT}
\end{figure}

Arranging the initial localized orbitals in \tesserae,
\begin{eqnarray}
  \label{EQ:phi0}
    \underline{\varphi}^{(0)} =
      \left(
         \underline{\varphi}^{(0)}_{A},
         \underline{\varphi}^{(0)}_{B},
         \ldots
      \right)
\,,
\end{eqnarray}
the approximation of local rotations for the localization step can be
written as
\begin{eqnarray}
  \label{EQ:locrot-a}
 \mbox{\it tessera}\,\,A&:& \hspace{5mm} 
  \underline{\varphi}^{L}_{A} =
        \sum_{C=1}^{N} \Delta^{LR}_{AC} \,\,
          \underline{\varphi}_C^{(0)}\,_0\underline{U}^{L}_{CA}
\,,           \\
  \label{EQ:locrot-b}
 \mbox{\it tessera}\,\,B&:& \hspace{5mm} 
  \underline{\varphi}^{L}_{B} =
        \sum_{C=1}^{N} \Delta^{LR}_{BC} \,\,
          \underline{\varphi}_C^{(0)}\,_0\underline{U}^{L}_{CB}
\,, \\ \nonumber
  & \ldots &
\end{eqnarray}
In Eq.~\ref{EQ:locrot-a}, $_0\underline{U}^{L}_{CA}$ is the $n_C
\times n_A$ block of $\underline{U}^{L}$ with the contributions of
the initial localized orbitals of \tessera\ $C$ to the target
localized orbitals of \tessera\ $A$. The column block made of all
$_0\underline{U}^{L}_{CA}$ with $\Delta^{LR}_{AC} = 1$ is
represented by the fill rectangular box shown in Fig.~\ref{FIG:LROT}.
This column block can be calculated approximately as the column
block of the submatrix of  $_0\underline{U}^{L}$ indicated in
Fig.~\ref{FIG:LROT} by the open square box. We can call this
submatrix $_0\underline{U}^{L}_{LRA}$, so that if
$\underline{\varphi}^{(0)}_{LRA}$ is a row vector with the initial
localized orbitals of all \tesserae\ $C$ having $\Delta^{LR}_{AC} =
1$ (of length \mbox{$n_{LRA} = \sum_{C=1}^{N} \Delta^{LR}_{AC} \,
n_C$}), then
\begin{eqnarray}
\label{EQ:ULRA}
  \underline{\varphi}^{ L }_{LRA} =
  \underline{\varphi}^{(0)}_{LRA} \,\, _0\underline{U}^{L}_{LRA}
\,.
\end{eqnarray}
$_0\underline{U}^{L}_{LRA}$ can be computed by simple application of
the localization method of choice
to the set of $n_{LRA}$ initial localized orbitals
$\underline{\varphi}^{(0)}_{LRA}$.
Note that only the $n_A$ orbitals of $\underline{\varphi}^{ L }_{LRA}$
corresponding to \tessera\ $A$ ($\underline{\varphi}^{ L }_A$)
are taken from Eq.~\ref{EQ:ULRA}.
Lowering the threshold of the local rotation table
improves the precision of this approximation systematically.

\mbox{}\\[-1cm]

The computation of the $_0\underline{U}^{L}_{LRA}$ matrices
(Eqs.~\ref{EQ:ULRA})
can be performed in parallel. This is graphically indicated in
the lower part of Fig.~\ref{FIG:LOOPS}.
The global scaling of this localization step is
$\sum_{B=1}^{N} {\cal O}(n_{LRB}^{\ell})$,
where $\ell$ is the order of the localization method of choice $L$
(e.g. 3 in Pipek-Mezey and 5 in Edmiston-Ruedenberg methods);
if the same number of orbitals are used in the local rotations
of all \tesserae, $\bar{n}_{LR}$,
the scaling is ${\cal O}(\bar{n}_{LR}^\ell N)$.

Since the
localization by local rotations among initial localized orbitals
goes together with the use of orbital-specific basis sets,
the resulting localized orbitals of \tessera\ $A$ must be represented
with the basis set $\underline{\chi}_A$ alone.
This can be achieved by truncation of
$\underline{\varphi}^{L}_{A}$ after the transformation
(Eq.~\ref{EQ:locrot-a}) or, alternatively,
by its projection on the $\underline{\chi}_A$ space,
$
  \underline{\chi}_A \, \left(
     \underline{\chi}^{\dagger}_A \, \underline{\chi}_A
                       \right)^{-1} \underline{\chi}^{\dagger}_A
 \,\,\underline{\varphi}^{L}_{A}
$.
We have not observed any practical advantage in projecting instead of
truncating, neither in the precision attainable in the total energy nor
in the convergence, in all the tests performed.

\subsection{Symmetry}

The orbitals resulting from the present method do not belong to
irreducible representations of the molecular point symmetry group
because they are not eigenfunctions of the totally symmetric
one-electron Hamiltonian $\hat{F}$. However, except for symmetry
breaking localization methods, they are related by the symmetry
operations of the molecule and this fact can be used to reduce
computing time.~\cite{PETERS:72}
In effect, all \tesserae\ orbitals can be obtained
by applying molecular symmetry operations, $\hat{R}$, to a list of
symmetry independent \tesserae, so that if $A$ is a symmetry
independent \tessera\ and \tessera\ $B$ is obtained from $A$ by
symmetry operation $\hat{R}$, then $\underline{\varphi}^{L}_{B} =
\hat{R}\,\underline{\varphi}^{L}_{A}$. In other words, the
diagonalizations and localizations described in
Sections~\ref{SEC:OSBS} and \ref{SEC:LR} can be performed only on
the list of symmetry independent \tesserae. The computation of
$\underline{F}^{L}_{A}$ (Eq.~\ref{EQ:FeffA}) can also profit from
this property.

Also, the site symmetry of the \tesserae\ can be used in
exactly the same way that molecular symmetry is used in standard
molecular calculations, because the matrix of the effective
Hamiltonian of any embedded  \tessera\ is blocked according to the
irreducible representations of the local point symmetry group of
that particular \tessera.

Besides the use of symmetry, one can also take advantage of
quasisymmetry for speeding up purposes. 
So, if two \tesserae\ $A$ and $B$ are  quasisymmetry related
by an operation $\hat{R}_q$ 
($\underline{\varphi}^{L}_{B} \approx
\hat{R}_q\,\underline{\varphi}^{L}_{A}$),
they can be treated as symmetry related for a number of macroiterations,
in which the list of \tesserae\ where the diagonalization and
localization steps are performed can be shortened,
finally releasing all quasisymmetry restrictions up to a full
convergence of the \mosaico\ calculation.

\subsection{Summary of the \mosaico\ algorithm}

In summary, the algorithm for a \mosaico\ calculation of a molecule
is the following:
\begin{enumerate}
\item \label{mosaico-step-1}
Take the current guess of localized orbitals of the molecule and
do, in parallel, for each symmetry independent \tessera:
  \begin{enumerate}
  \item
  compute the embedded \tessera\ effective Hamiltonian matrix
  (Eqs.~\ref{EQ:FeffA}, \ref{EQ:rhoHrho-ok}, and \ref{EQ:thirdterm}),
  \item
  diagonalize it (Eq.~\ref{EQ:a}) and
  compute new \tessera\ orbitals (Eq.~\ref{EQ:neworb}),
  \end{enumerate}
\item \label{mosaico-step-2}
Take all the orbitals produced in step \ref{mosaico-step-1} and
do, in parallel, for each symmetry independent \tessera:
  \begin{enumerate}
  \item
  compute the $_0\underline{U}^{L}_{LRA}$
  unitary matrix (Eq.~\ref{EQ:ULRA})
  corresponding to the localization method of choice $L$
  by applying the corresponding localization algorithm,
  and take the  columns that correspond to the current \tessera,
  \item
  compute the target \tessera\ localized orbitals (Eq.~\ref{EQ:locrot-a}),
  \end{enumerate}
\item
Check for convergence and iterate on step~\ref{mosaico-step-1} if
necessary. Compute properties upon convergence.
\end{enumerate}
\label{EQ:project}

The parallel steps \ref{mosaico-step-1} and \ref{mosaico-step-2}
are schematically represented in Fig.~\ref{FIG:LOOPS}.
\section{RESULTS}
\label{SEC:Results}

In this section, we present the results of monitoring calculations
on poly(ethylene oxide) molecules, \peo, (Sec.~\ref{SEC:PEO}) 
and three dimensional CO clusters, (CO)$_m$, (Sec.~\ref{SEC:CO}) 
aimed at showing the convergence of the parallel calculations to the 
right solutions,
the convergence of the total energies with the size of the orbital-specific
basis sets towards the exact values,
and the linear-scaling of the method.
We also include embedded cluster calculations on defective systems
resulting from a chemical substitution of one O atom by a S atom in 
poly(ethylene oxide), (Sec.~\ref{SEC:EC}) aimed at showing the performance
of embedded cluster calculations vs. full system calculations.
 
All the calculations
are single point energy calculations with an Extended H\"{u}ckel
(EH) Hamiltonian.~\cite{HOFFMANN:63} This is a convenient choice to
monitor the \mosaico\ procedure because, on the one hand, the
Hamiltonian and overlap matrices of the EH method are isomorphous
with their \abinitio\ counterparts, and on the other, the
Hamiltonian is not self-consistent and its computation is
straightforward. In this way, the analysis of timing and scaling 
is focused on the orbital optimization or diagonalization part and
free from contaminations due to the computation of the Hamiltonian
matrix.
The calculations have been performed with a 
2GB RAM personal computer with the program
Mosaico;~\cite{MOSAICO} the Extended H\"{u}ckel Hamiltonian and
overlap matrices have been calculated with the program
EHT.~\cite{EHT} 
Although the loops performed mimic parallel loops (Fig.~\ref{FIG:LOOPS}),
the total elapsed times shown hereafter correspond to sequential loops 
performed in a single processor.
In the present version of the program
we have paid special attention to the scaling features,
whereas the prefactors are highly improvable.
This fact,
together with the average performance of the personal computer used, 
makes the absolute values of the elapsed times shown in this section
of little value; instead,
it is the scaling of the method what is relevant.

\subsection{Poly(ethylene oxide)}
\label{SEC:PEO}

Poly(ethylene oxide) (PEO)~\cite{GRAY:91} is a polymer widely used
in the field of polymer electrolytes of molecular formula \peo. The
molecular calculations we present for different numbers of monomers,
$m$, use the localization method of projected localized molecular
orbitals (PLMO)~\cite{RUEDENBERG:82} (see Appendix) using a very
simple set of reference orbitals: $s_A + s_B$ for all A-B pairs of
bonded atoms, plus $p_y(O) + p_z(O)$ and $p_y(O) - p_z(O)$ for all
oxygen atoms, where $y$ and $z$ are local cartesian axis on the
oxygen assuming the C-O-C atoms define a $xy$ plane with the $y$
axis bisecting the C-O-C angle.

In all these calculation we used the following definition for the
\tesserae\ or subsystems:
One \tessera\ made of 10 orbitals localized in the sigma bonds and
lone pairs of CH$_3$OCH$_2$--
[which, in the localization method used for these particular calculations,
are the 10 orbitals with one-to-one maximum overlap with the reference
orbitals
$s(H_1)+s(C_1)$,
$s(C_1)+s(H_2)$,
$s(C_1)+s(H_3)$,
$s(C_1)+s(O_1)$,
$p_y(O_1)+p_z(O_1)$,
$p_y(O_1)-p_z(O_1)$,
$s(O_1)+s(C_2)$,
$s(C_2)+s(H_4)$,
$s(C_2)+s(H_5)$,
$s(C_2)+s(C_3)$],
plus $m-2$ \tesserae\ made of 9 orbitals localized in the sigma bonds and
lone pairs of the next CH$_2$OCH$_2$-- groups,
and a final \tessera\ of 9 orbitals localized in the sigma bonds and
lone pairs of the terminal CH$_2$OCH$_3$.

\begin{figure}
  \begin{center}
  \mbox{\resizebox{7cm}{!}{\includegraphics*{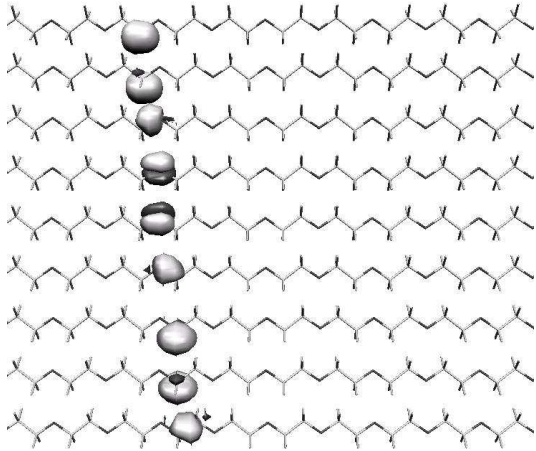}}}
  \end{center}
  \mbox{}\\[-24pt]
  \caption{
           The nine localized orbitals that constitute
           one of the bulk \tesserae\ used in H(CH$_2$OCH$_2$)$_{30}$H.
          }
  \label{FIG:PEO-TESS9}
\end{figure}

We performed the \mosaico\ calculations using three different
orbital-specific basis sets: In the calculations labeled 1N, the
orbitals of each \tessera\ have been represented with a subset of
the global basis set consisting of all the basis set functions of
the atoms involved in the bonds and lone pairs of the \tessera\ plus
those of the atoms involved in bonds and lone pairs of the first
neighbor \tesserae\ or monomers. In the 2N and 3N calculations, the
orbital-specific basis sets were extended to second and third
neighbor monomers.
All the calculations converge to the same results regardless of the
initial guess and the iteration procedure (parallel or any kind of
sequential choice).
The 9 localized orbitals which constitute one of the
the bulk \tesserae\ of H(CH$_2$OCH$_2$)$_{30}$H
are shown in Fig.~\ref{FIG:PEO-TESS9}.

\begin{figure}[pt]
  \begin{center}
  \mbox{\resizebox{8.5cm}{!}{\includegraphics*{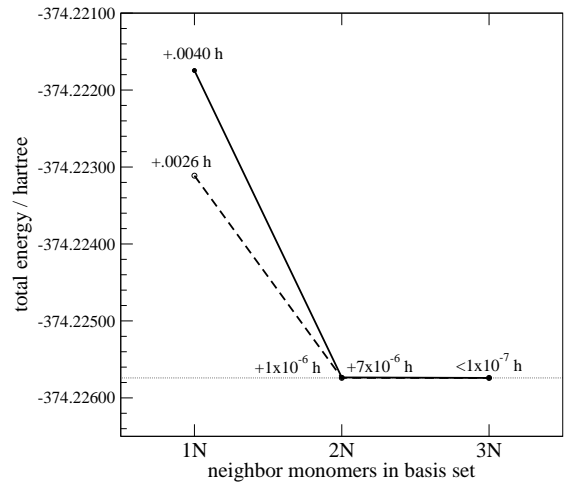}}}
  \end{center}
  \mbox{}\\[-0.5cm]
  \caption{
           Total energy of H(CH$_2$OCH$_2$)$_{30}$H
           as a function of the orbital-specific basis set size.
           1N, 2N, and 3N labels indicate that each \tessera\ is
           calculated with the basis set functions of all atoms up to
           first, second, and third neighbor monomers, respectively.
           Energy losses with respect to the exact canonical energy
           are indicated on the lines, in hartree units.
           Full line: correct calculation of the total energy.
           Dashed line: total energy calculated
           under the assumption of perfect orthogonality between
           the localized orbitals.
          }
  \label{FIG:PEO-ETOT}
\end{figure}

\begin{figure}
  \begin{center}
  \mbox{\resizebox{8.5cm}{!}{\includegraphics*{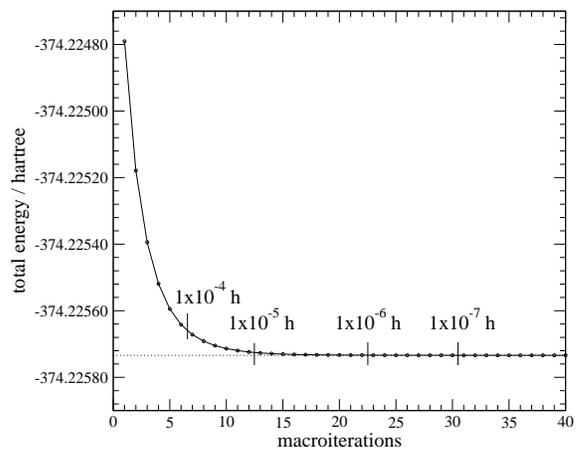}}}
  \end{center}
  \caption{
           Convergence with macroiterations of the total energy of
           H(CH$_2$OCH$_2$)$_{30}$H with 2N orbital-specific basis set.
           Convergence to several subunits of hartree are indicated.
           A similar number of macroiterations for convergence has
           been found in all poly(ethylen oxide) polymers studied.
          }
  \label{FIG:PEO-EvsIter}
\end{figure}

Fig.~\ref{FIG:PEO-ETOT} shows
the total energy of H(CH$_2$OCH$_2$)$_{30}$H
as a function of the orbital-specific basis sets used,
which converges to the exact energy in the limit of a standard
calculation where all orbitals are spanned in a common basis set.
As the size of the OSBS increases, the lack of full orthogonality
between orbitals of differente \tesserae\ originated by the
basis set truncation becomes negligible and, accordingly,
the difference between the total energy properly computed and
the total energy computed under the assumption that the orbitals
are fully orthogonal vanishes.
The energy loss per monomer
due to the use of orbital-specific basis sets instead of a common
basis set for all orbitals,
is presented in Table~\ref{TAB:ONE}
for several orbital-specific basis sets
as a function of the molecular size.
The predictability of the  energy losses is apparent.
Fig.~\ref{FIG:PEO-EvsIter} shows the convergence with macroiterations
of the total energy of the H(CH$_2$OCH$_2$)$_{30}$H polymer at the 2N level
of orbital-specific basis set.

\begin{table}[pb]
\caption{
         Energy loss per monomer with respect to canonical calculations, 
         in hartree/monomer,
         of poly(ethylen oxide) H(CH$_2$OCH$_2$)$_{m}$H molecules
         and (CO)$_{m}$ clusters,
         as calculated with several orbital-specific basis sets.
}
\label{TAB:ONE}
    \renewcommand{\arraystretch}{0.64}   
\begin{ruledtabular}
\begin{tabular}{ ccrrr }
\\
  & OSBS 
  & \multicolumn{3}{c}{ ${(E-E_{canonical})}/{m}$ }
\\
  \cline{2-2} \cline{3-5} 
\\
    \multicolumn{2}{l}{ H(CH$_2$OCH$_2$)$_m$H }
  & \multicolumn{1}{c}{$m=10$ }
  & \multicolumn{1}{c}{$m=20$ }   
  & \multicolumn{1}{c}{$m=50$ }  
\\
              \cline{3-3} \cline{4-4} \cline{5-5} 
\\
  & 1N 
  & $1.14 \times 10^{-4}$ 
  & $1.28 \times 10^{-4}$ 
  & $1.37 \times 10^{-4}$ 
\\
  & 2N 
  & $1.74 \times 10^{-7}$ 
  & $2.09 \times 10^{-7}$ 
  & $2.30 \times 10^{-7}$ 
\\
  & 3N 
  & $< 1 \times 10^{-11}$ 
  & $< 1 \times 10^{-11}$ 
  & $2.0 \times 10^{-10}$ 
\\
\\
    \multicolumn{1}{c}{ (CO)$_m$ }
  &
  & \multicolumn{1}{c}{$m=13$ }
  & \multicolumn{1}{c}{$m=63$ }   
\\
              \cline{3-3} \cline{4-4} 
\\
  & 1N 
  & $3.56 \times 10^{-7}$ 
  & $7.74 \times 10^{-7}$ 
\\
  & 2N 
  & $2.34 \times 10^{-7}$ 
  & $5.13 \times 10^{-7}$ 
\\
  & 3N 
  & $3.3 \times 10^{-8}$ 
  & $3.7 \times 10^{-8}$ 
\\
\end{tabular}
\end{ruledtabular}
\end{table}

\begin{figure}
  \begin{center}
  \mbox{}\\[-1.8cm]
  \mbox{\resizebox{8.0cm}{!}{\includegraphics*{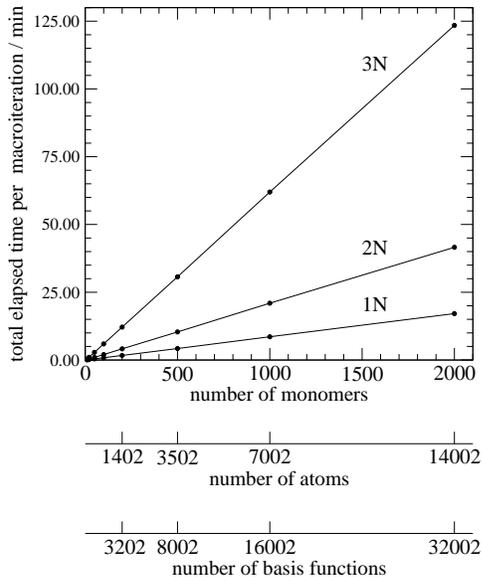}}}
  \end{center}
  \mbox{}\\[-1.5cm]
  \caption{
           Total wall clock elapsed times per macroiteration
           in the calculation of poly(ethylen oxide) molecule
           H(CH$_2$OCH$_2$)$_{m}$H
           ($m=10, 20, 50, 100, 200, 500, 1000, 2000$).
          }
  \label{FIG:PEO-TIME}
\end{figure}

\begin{figure}
  \begin{center}
  \mbox{\resizebox{8.0cm}{!}{\includegraphics*{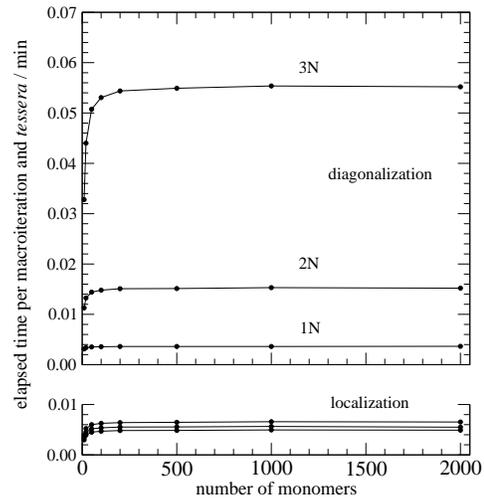}}}
  \end{center}
  \mbox{}\\[-2.5cm]
  \caption{
           Elapsed times per macroiteration and \tessera\
           in the calculation of poly(ethylen oxide) molecule
           H(CH$_2$OCH$_2$)$_{m}$H.
           Up: Diagonalization step. Down: Localization step.
          }
  \label{FIG:PEO-TESSERATIME}
\end{figure}

Fig.~\ref{FIG:PEO-TIME} shows the wall clock elapsed time per
macroiteration in the calculation of the H(CH$_2$OCH$_2$)$_{m}$H
molecules as a function of the number of monomers $m$ (which in this
case coincides with the number of \tesserae, $N$), of atoms, and of
basis set functions. 
%
The times per macroiteration and \tessera\
spent in the diagonalization step and in the localization step
are shown in Fig.~\ref{FIG:PEO-TESSERATIME}.
The horizontal lines reflect the \order($N$) scalings of both steps.
The \order($\langle {b}^3 \rangle$) scaling of the diagonalization step 
is clearly
reflected in the separation between the 1N, 2N, and 3N horizontal lines.
The drop of the lines at low number of monomers is
due to surface effects:
The ratio between edge \tesserae\ and bulk \tesserae\ is significant in
small polymers and,
since edge \tesserae\ are less demanding in terms of
orbital-specific basis set and
in terms of number of inter-\tesserae\ interactions,
they reduce the computing time with respect to what it would be
if all of them were bulk \tesserae.
As we will see below, this fortunate surface effect,
which lowers time with respect to a set of $N$ bulk \tesserae,
is much more pronounced in 3D systems.
The localization times shown in the bottom part of
Fig.~\ref{FIG:PEO-TESSERATIME} are very similar 
in the 1N, 2N and 3N calculations.
This is so because they depend basically on the number of orbitals used
in the local rotations, which is the same in the three calculations.
For the particular localization method we used for these calculation, 
PLMO, the elapsed times scale as
$\sum_{B=1}^{N}$\order(${n}_{LRB}^{3}$)
$\approx$~\order($\langle{n}_{LRB}^{3}\rangle$)\order($N$). 
The small dependence with the size
of the orbital-specific basis sets is related with the lengths of
the matrix transformations in Eq.~\ref{EQ:PLOLR}.

\subsection{(CO)$_m$ clusters}
\label{SEC:CO}

\begin{figure}[pt]
  \begin{center}
  \mbox{}\\[-12pt]
  \mbox{\resizebox{8.0cm}{!}{\includegraphics*{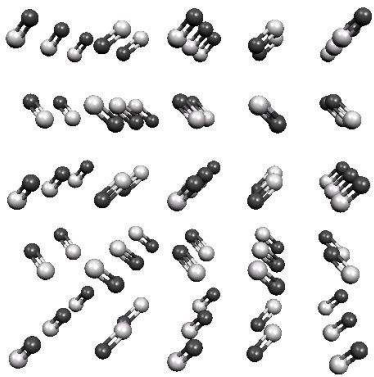}}}
  \end{center}
  \mbox{}\\[-48pt]
  \caption{
           (CO)$_{63}$ cluster as a piece of the
           $\alpha$ phase of solid carbon monoxide (space group $P2_{1}3$).
          }
  \label{FIG:CO63}
\end{figure}

We performed \mosaico\ calculations on three dimensional
(CO)$_{m}$ clusters of several sizes,
extracted from the  crystalline structure of the $\alpha$ phase of
solid carbon monoxide
($P2_{1}3$ spatial group, CO bond length $r$(C-O)~=~1.128~\AA,
cell constant $a_0$~=~5.64~\AA ),~\cite{HALL:76}
which is an interesting material 
known to experience irreversible
photopolymerization under pressure.~\cite{KATZ:84,BERNARD:98}
In this crystal,
a bulk CO molecule has a first-neigbhor coordination number 12
(CO molecules at a distance between centers of gravity of 3.99~\AA),
and second- and third-neigbhor coordination numbers 18 and 42 
(CO molecules at 5.64~\AA\ and 6.91~\AA, respectively).
The (CO)$_{63}$ cluster is represented in Fig.~\ref{FIG:CO63}
as an example.

In these calculations we also used the PLMO localization method, with
the reference orbitals defined as the $5 m$ valence occupied
canonical orbitals of the $m$ isolated CO molecules. We defined each
\tessera\ to be made of 5 orbitals localized in the spatial region
of a CO molecule, which, for the chosen localization method, means
the 5 localized orbitals with maximum overlap with the canonical
orbitals of the CO molecule.
The orbital-specific basis sets used have been labeled 1N, 2N, and 3N
when the basis set of a \tessera\ consists of the basis set functions of its
C and O atoms plus the basis set functions of the atoms of the
first, second, and third-neighbor CO molecules, respectively (see above).

\begin{figure}
  \begin{center}
  \mbox{}\\[-0.8cm]
  \mbox{\resizebox{8.0cm}{!}{\includegraphics*{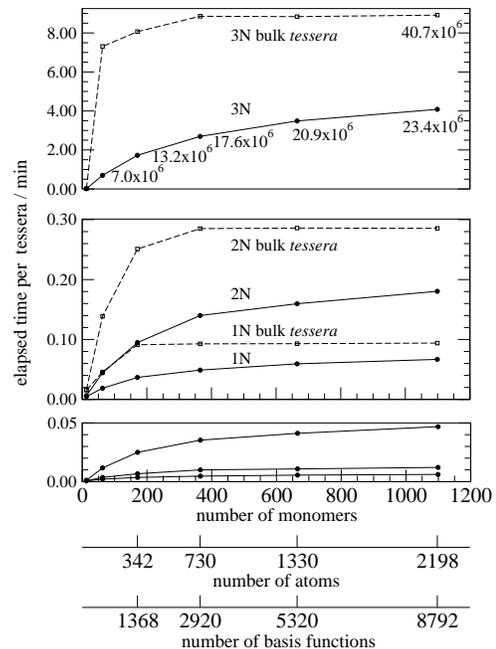}}}
  \end{center}
  \caption{
           Elapsed times per macroiteration and \tessera\
           spent in the calculation of (CO)$_{m}$ clusters.
           Down: Localization step of the 1N, 2N, and 3N OSBS calculations.
           Middle: Diagonalization step of the 1N and 2N OSBS calculations.
           Up: Diagonalization step of the 3N OSBS calculations;
           The values of $\langle b^3 \rangle = (\sum_{A=1}^{m} b_A^3)/m$,
           where $b_A$ is the number of basis set functions in the OSBS
           of \tessera\ A, are indicated.
          }
  \label{FIG:CO-TESSERATIME}
\end{figure}

The energy losses per CO molecule (Table~\ref{TAB:ONE}) are small
and diminish as the size of the OSBS increases.
Times per macroiteration and \tessera\
spent in the diagonalization step and in the localization step
are shown in Fig.~\ref{FIG:CO-TESSERATIME}.
The diagonalization times spent in a inner or bulk \tessera,
which is the most demaning, are included as a reference.
This time is constant for all clusters where
the inner \tessera\ is a truly bulk \tessera,
that is, where its OSBS and its interaction tables die off before
the limits of the cluster;
this situation is reached in a smaller cluster for the 1N OSBS 
and in a bigger cluster for the 3N OSBS.
Naturally,
the times spent per \tessera\ in the full system calculations are
lower than the times spent in the bulk \tesserae,
the differences showing the importance of surface effects:
smaller clusters and bigger orbital-specific basis sets have a larger
ratio of surface/bulk \tesserae\ and, accordingly, a larger time reduction
with respect to the bulk \tesserae.
In a large $m$ regime, the bulk \tesserae\ are dominant and set the
assymptotic limit of the full system times, which scale linearly
with the cluster size.
The \order($\langle b^3 \rangle$) dependence is shown by the 
asymptotic values of the 1N, 2N, and 3N lines,
as well as by the values of $\langle b^3 \rangle = (\sum_{A=1}^{m} b_A^3)/m$
printed along the 3N line in Fig.~\ref{FIG:CO-TESSERATIME}.
The localization times (bottom of Fig.~\ref{FIG:CO-TESSERATIME})
are very much independent of the size of the OSBS because they
are directly dependent in the number of occupied orbitals included
in the local rotations, which is the same in 1N, 2N, and 3N calculations;
the small dependence shown in the Figure is due to the fact that 
the transformations down to the basis set level depend on the OSBS size.

\subsection{Embedded cluster calculations}
\label{SEC:EC}

The \mosaico\ method can be used for embedded cluster calculations,
where the computational effort is focused on an active site of a molecule,
comprising only a number of relevant \tesserae,
while the rest of it is taken from a previous calculation on a similar
molecule and frozen.
In this section we show the results of embedded cluster calculations
on 
H(CH$_2$OCH$_2$)$_{p}$-CH$_2$SCH$_2$-(CH$_2$OCH$_2$)$_{p}$H.

\begin{figure}[pb]
  \begin{center}
  \mbox{}\\[-12pt]
  \mbox{\resizebox{8.0cm}{!}{\includegraphics*{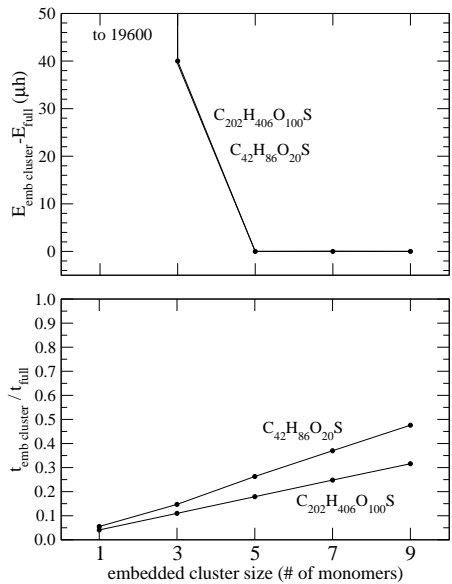}}}
  \end{center}
  \mbox{}\\[-1.5cm]
  \caption{
           Embedded cluster calculations on
           H(CH$_2$OCH$_2$)$_{p}$-CH$_2$SCH$_2$-(CH$_2$OCH$_2$)$_{p}$H
           ($p=10, 50$)
           with a 5N OSBS for the variational \tessera\
           and 3N OSBS for the remaining \tesserae.
           The abscissa labels correspond to the number of
           variational \tesserae\ included in the active clusters;
           the defective \tessera\ is always the central one.
           Above:
           Total energy errors, in $\mu$hartree,
           with respect to full molecule \mosaico\ calculations
           using 5N OSBS in the 11 central \tesserae\ and
           3N OSBS in the rest;
           the result of the smallest cluster is indicated.
           Below:
           Elapsed time per macroiteration,
           as a fraction of the time per macroiteration
           of the full molecule \mosaico\ calculation.
          }
  \label{FIG:PEO-S-EC}
\end{figure}

H(CH$_2$OCH$_2$)$_{p}$-CH$_2$SCH$_2$-(CH$_2$OCH$_2$)$_{p}$H
can be regarded as the result of creating a chemical defect
in H(CH$_2$OCH$_2$)$_{m}$H, with $p=(m-1)/2$,
by substitution of the central oxygen by a sulfur atom.
We may expect the localized orbitals distant from the S atom
in the ``defective" molecule to be very similar to the orbitals
of the ``perfect" molecule localized in the same spatial regions
and, accordingly, we may take them from a previous calculation
on H(CH$_2$OCH$_2$)$_{m}$H and use them in a \mosaico\ calculation of
H(CH$_2$OCH$_2$)$_{p}$-CH$_2$SCH$_2$-(CH$_2$OCH$_2$)$_{p}$H
where they are kept frozen;
this defines an embedded cluster \mosaico\ calculation.
Canonical and \mosaico\ calculations on the $m=21$ polymer
(using the same nuclear configurations
before and after the creation of the S defect)
reveal that the precision reached with a 3N orbital-specific basis set on the
perfect polymer requires a better OSBS after creating this chemical defect:
A 5N basis set for the central defective \tessera\ and its 5 neighbor 
\tesserae\,
together with a 3N basis set for the remaining \tesserae,
gives an energy error of $4.5 \times 10^{-9}$ hartree/monomer.
Taking this into account,
we performed embedded cluster calculations using a 5N OSBS for the 
variational \tesserae\ and 
taking the orbitals that remain frozen for the rest of \tesserae\
in the polymer
from the 3N OSBS calculation on the ``perfect" polymer molecule
H(CH$_2$OCH$_2$)$_{m}$H.
The total energy errors of these embedded cluster calculations
(with respect to the full molecular \mosaico\ calculations)
on the polymers of 21 and 201 monomers 
are shown in Fig.~\ref{FIG:PEO-S-EC} as a function of 
the active cluster size (number of active \tesserae).

It is shown that the errors are too large to be acceptable if only
the central defective \tessera\ is active, whereas they drop to 
an acceptable value of $40 \times 10^{-6}$~hartree when the orbitals
of the first-neigbhor \tesserae\ are optimized as well,
and to less than $1 \times 10^{-6}$~hartree when second-neigbhor
\tesserae\ are part of the variational embedded cluster.
These errors are the same in the two $m=21$ and $m=201$ polymers,
as corresponds to the local nature of the chemical defect.
The times spent in the embedded cluster calculations, as a fraction of 
the times spent in the respective full system calculations, are shwon
at the bottom of Fig.~\ref{FIG:PEO-S-EC}.
Overall, this figure illustrates the potentiallity of 
embedded cluster calculations where the transferability of 
the localized orbitals of a localization method of choice is exploited.

\section{CONCLUSIONS}
\label{SEC:Conclusions}

We presented
a new linear-scaling method for the energy minimization step
of semiempirical and first-principles Hartree-Fock and Kohn-Sham calculations,
which we abbreviated under the name \mosaico.
In this method,
a set of embedded \tessera\ pseudoeigenvalue coupled equations
is solved in a  building-block self-consistent fashion,
which results in optimum occupied localized orbitals of 
any localization method of choice,
represented with orbital-specific basis sets.
The \mosaico\ method is parallel at a high level of the calculation.
It can be used in full system calculations 
as well as in embedded cluster calculations, 
where only an active fraction of the localized molecular orbitals 
of the whole system are variational while the rest are taken from
a similar molecule and kept frozen.

We presented
the results of monitoring, single point energy calculations 
with the extended H\"uckel Hamiltonian 
on poly(ethylene oxide) molecules and
three dimensional carbon monoxide clusters with very large number
of basis set functions.
Total energy losses due to the use of orbital-specific basis sets
are small for reasonably small sizes of these and
total energies converge to the canonical values
when the orbital-specific basis sets are increased towards the
limit of a common basis set for all the localized orbitals.
Convergence of total energy with self-consistent macroiterations
is good and elapsed times per macroiterations have been shown to
scale linearly with the molecular size.
Besides the full system calculations,
the good performance of the much less demanding embedded cluster approach 
has been shown in total energy calculations on 
defective systems resulting from chemical substitution of an oxygen
atom by a sulfur atom in poly(ethylene oxide) molecules.
The transferability of the localized orbitals of a given localization method 
between similar molecules has been shown to lead to the same total energy 
precision than full molecular calculations at a fraction of the 
computational cost.

\acknowledgments
    This research was supported in part by 
    Direcci\'on General de Investigaci\'on,
    Ministerio de Ciencia y Tecnolog\'{\i}a, Spain,
    under contract BQU2002-01316,
    and Ministerio de Educaci\'on, Cultura y Deportes, Spain,
    Accciones de Movilidad PR2003-0024 and PR2003-0027.
    We are very grateful to Professor Martin Head-Gordon,
    University of California, Berkeley, and to his research group
    for their hospitality.
\section*{APPENDIX}
\subsection{Projected localized molecular orbitals}

A simple localization method has been proposed by 
Ruedenberg \etal,~\cite{RUEDENBERG:82}
which has not reached the popularity of other methods like 
Boys,~\cite{BOYS:60} 
Edmiston-Ruedenberg,~\cite{EDMISTON:63} 
and Pipek-Mezey~\cite{PIPEK:89} methods.
Although it has been formulated in a context of atoms-in-molecules,
it is a general method. 
Here, we reformulate it in the general case and in the case of 
localization by local rotations (Sec.~\ref{SEC:LR}), 
using the present notation.
 
Given a set of $n$ input occupied (canonical or localized) orthogonal orbitals
of a molecule,
$\underline{\varphi}^{(0)} =
 \left( \mid \varphi^{(0)}_1 \rangle,
        \mid \varphi^{(0)}_2 \rangle,
        \ldots,
        \mid \varphi^{(0)}_n \rangle
 \right)
$,
and a set of $n$ reference arbitrary orbitals,
$\underline{\xi} =
 \left( \mid \xi_1 \rangle,
        \mid \xi_2 \rangle,
        \ldots
        \mid \xi_n \rangle
 \right)
$,
the $n$ projected localized orthogonal orbitals,
$\underline{\varphi}^{PLMO} =
 \left( \mid \varphi^{PLMO}_1 \rangle,
        \mid \varphi^{PLMO}_2 \rangle,
        \ldots,
        \mid \varphi^{PLMO}_n \rangle 
 \right)
$,
are defined as those resulting from a unitary transformation
of $\underline{\varphi}^{(0)}$ which are most similar to $\underline{\xi}$.
The PLMOs correspond to maximizing the one-to-one overlaps with the
reference orbitals $\underline{\xi}$, or the functional
$\sum_{i=1}^{n}\mid\langle \varphi^{PLMO}_i \mid \xi_i \rangle\mid$,
under orthogonality constraints~\cite{FRANCISCO:86}
and can be computed as~\cite{RUEDENBERG:82,FRANCISCO:86}
\begin{eqnarray}
  \underline{\varphi}^{PLMO} 
  = 
  \hat{\rho}\,\underline{\xi}\,
    \left[
    \underline{\xi}^{\dagger}\,\hat{\rho}\,\underline{\xi}
    \right]^{-\frac{1}{2}}
  = 
  \underline{\varphi}^{(0)}\, _0\underline{U}^{PLMO}
\,,
\end{eqnarray}
with
\begin{eqnarray}
  _0\underline{U}^{PLMO}
  = 
    \left( \underline{\varphi}^{(0) \dagger}\underline{\xi} \right)
    \left[
      \left( \underline{\varphi}^{(0) \dagger}\underline{\xi} \right)^\dagger
      \left( \underline{\varphi}^{(0) \dagger}\underline{\xi} \right)
    \right]^{-\frac{1}{2}}
\,.
\end{eqnarray}
In the approximation of local rotations (Section~\ref{SEC:LR}),
the projected localized orbitals of a \tessera\ $A$
are computed out of $n_{LRA}$ ($n_{LRA} > n_{A}$)
input localized orbitals $\underline{\varphi}^{(0)}_{LRA}$
and $n_{LRA}$ reference orbitals $\underline{\xi}_{LRA}$,
both of them including the localized/reference orbitals belonging
to the \tesserae\ included in the local rotations,
with the expression:
\begin{eqnarray}
  \underline{\varphi}^{PLMO}_A
  = 
  \underline{\varphi}^{(0)}_{LRA}\, _0\underline{U}^{PLMO}_{LRA}
\,,
\end{eqnarray}
being
\begin{eqnarray}
\label{EQ:PLOLR}
  && _0\underline{U}^{PLMO}_{LRA}
  = 
  \Big\{
    \left( \underline{\varphi}^{(0) \dagger}_{LRA}
           \,\underline{\xi}_{LRA} \right)
 \times \hfill
 \nonumber \\
  &&  \left[
      \left( \underline{\varphi}^{(0) \dagger}_{LRA}
             \,\underline{\xi}_{LRA} \right)^\dagger
      \left( \underline{\varphi}^{(0) \dagger}_{LRA}
             \,\underline{\xi}_{LRA} \right)
    \right]^{-\frac{1}{2}}
  \Big\}_{n_A-\mbox{col}}
\,,
\end{eqnarray}
where it has been indicated that only the $n_A$ columns 
that correspond to  \tessera\ $A$
are computed and used.

Among the advantages of the PLMO localization method are
its speed and its simplicity,
because the usual iterative optimization 
procedures involved in localization~\cite{BOYS:60,EDMISTON:63,PIPEK:89}
are substituted by a one-step calculation of
the reciprocal square root of a symmetrical matrix,
which is an \order($n^3$) process
(or \order($n_{LRA}^3$) in local rotations).
Its main disadvantage is the requirement of an external,
arbitrary set of reference orbitals, $\underline{\xi}$.
Although this is a limitation in calculations of reactivity,
it is not a practical problem in molecular structure calculations
where the nature of the bonds is known in advance and finding
good reference bond and lone pair orbitals is not difficult.
We may remark that
the application of the PLMO method with a reference consisting
of a given set of localized orbitals, e.g. Edmiston-Ruedenberg's,
leads exactly to that set of orbitals.
This property can be exploited in many ways
and, in particular, in order to remove the arbitrariness inherent to
the PLMO method. 
For instance, it can be used to produce reference orbitals 
for the PLMO method (to be used in large molecules)
out of Edmiston-Ruedenberg's or other non-arbitrary localized sets
computed in selected sets of relatively small molecules.
Also, a \mosaico\ calculation addressed to produce a given set 
of localized orbitals like Edmiston-Ruedenberg's can be safely performed
using the chosen localization method in some macroiterations and
the faster PLMO method, with the current ER orbitals as a reference set,
in the rest of them.


\begin{thebibliography}{50}
\bibitem[{MOS()}]{MOSAICO}
\bibinfo{note}{Mosaico 1.0 is a program for building-block and embedding
  calculations based on the use of localized orbitals and orbital-specific
  basis sets, written by L. Seijo, Departamento de Qu\'{\i}mica, Universidad
  Aut\'onoma de Madrid, Madrid, Spain.}
\end{thebibliography}



\end{document}